# Development of a Scale to Measure Technology Acceptance in Smart Agriculture

Rosemary J Thomas[1] · Rebecca Whetton[2] · Andy Doyle[3] · David Coyle[4]

**Abstract** This paper describes the development of a scale to measure technology acceptance in smart agriculture. The scale is intended for use in diverse situations, ranging for the evaluation of existing technologies already in widespread use, to the evaluation of prototype systems. A systematic screening of prior literature was conducted to identify initial scale items regarding how technology acceptance is currently understood and measured. These items were iteratively reviewed and systematically categorised to develop the final scale proposed in this paper. In future work, this scale will be validated through user studies. The purpose of this paper is to make the initial scale available to the research community with a view to initial use and further evaluation.

**Funding**
This research was undertaken as part of the CONSUS project and is funded under the Science Foundation Ireland Strategic Partnerships Programme (16/SPP/3296) and is co-funded by Origin Enterprises Plc.

**Conflicts of interest/Competing interests**
Not applicable

[1] R. J. Thomas
University College Dublin, School of Computer Science, Dublin, Ireland
ORCID 0000-0001-7998-4476, E-mail: rosemaryjthomas@acm.org
[2] R. Whetton
University College Dublin, School of Biosystems and Food Engineering, Dublin, Ireland
E-mail: rebecca.whetton@ucd.ie
[3] A. Doyle
The Irish Farmers Journal, Tillage Editor, Dublin, Ireland
E-mail: adoyle@farmersjournal.ie
[4] D. Coyle
University College Dublin, School of Computer Science, Dublin, Ireland
ORCID 0000-0002-5103-0379, E-mail: d.coyle@ucd.ie

## Introduction

Increasing populations and climate change have had a significant impact on global agriculture, emphasising the need for both increased food security and greater sustainability. The last decade has also witnessed an explosion of information and communication technologies (ICTs) that may significantly disrupt traditional agriculture, both positively and negatively. Smart agriculture has the potential to help in addressing global agricultural challenges, but success will be dependent on the degree to which new technologies are accepted by the agriculture community. It is therefore important to measure the acceptance of such technology. However, such measurement is difficult as no validated and established measure of technology acceptance in smart agriculture currently exists.

Nadal et al. (2020) define acceptability from the perspective of the user i.e., the process that the user goes through in decision making to make sure that an object is suitable and accepted. A number of formal scales and models currently exist to assess acceptance. For example, the Technology Acceptance Model (TAM) has been validated in different domains such as IT usefulness and ease of use (Davis 1989) and IT adoption by employees (Venkatesh and Bala 2008). King and He (2006) studied the meta-analysis of 88 TAM studies in different areas in Information Technology. However, no such scale exists within the domain of smart agriculture. The purpose of this paper is to develop such a scale.

Our scale builds on the Theoretical Model of Acceptance or Theoretical Framework of Acceptability (TFA). According to TFA, to measure technology acceptance 7 dimensions of acceptability (Sekhon, Cartwright, and Francis 2017) are used: Affective Attitude, Burden, Ethicality, Intervention/System Coherence, Opportunity Costs, Perceived Effectiveness, and Self Efficacy. All these dimensions look at acceptability from the perspective of the user rather than the focal point of the intervention or system. This is further discussed in the 'Literature Screening' section.

When we investigated the literature in crop farming in smart agriculture, we found out that there is no standardised method or process to evaluate technology acceptance that has undergone proper validation and most researchers have so far had to use their own measures without proper validation. Therefore, we would like to develop a reliable scale that is user-focused, that includes multiple factors as sub-scale items, and each sub-scale contains multiple items covering all the 7 dimensions of acceptability.

This paper provides a detailed description of the development of the technology acceptance scale based on the Theoretical Acceptance Model in smart agriculture. We collected 198 scale items from the 16 peer-reviewed papers that were shortlisted through a systematic literature screening process. The 198 items were reduced to 47 scale items after rigorous reviews. This is further discussed in the 'Methodology' section.

This paper contribution to the smart agriculture community by presenting the full procedure to develop a standardised, domain specific technology acceptance scale that is user-focused and would help close the gap in the present literature of unstandardised and unvalidated scales.

## Literature Screening

Papers were identified by performing an online search of journal articles and conference proceedings from the following databases: ACM Digital Library, Scopus, CABI and Business Source Complete (EBSCOhost). CABI database was included as it is dedicated to agriculture and the EBSCOhost database was included as it covered topics in agriculture. Both Scopus and ACM Digital Library included results from IEEE Xplore using the keywords from Table 2. The PICO (Population Intervention Comparison and Outcome) acronym (Methley et al., 2014) was adapted to describe the inclusion and exclusion criteria used for the identification of papers. Our focus was on the user studies in crop farming based on technology interventions (see Table 1).

*Table 1: Eligibility criteria*

|  | **Inclusion Criteria** | **Exclusion Criteria** |
|---|---|---|
| **POPULATION** | Minimum age of 18 years<br>Participants who have or had a farming or agricultural (crops) background | Children<br>Not farming or agricultural (crops) related |
| **INTERVENTION TYPE** | Technology-enabled interventions (e.g., apps, websites, tools that assess crop farming technology acceptance) used by the participants<br>Technology-enabled apps, e.g., interventions and their impact on such as soil improvement or crop production<br>Online resources that improve crop yields | Diagnostics technology-enabled interventions (e.g., soil sampling)<br>Technology-enabled interventions not directly used by farmers or agronomists |
| **STUDY TYPE** | Qualitative and quantitative studies exploring the technology acceptability which includes effectiveness, usability, user experience and design of a technology-based intervention (i.e., RCTs (Randomized Controlled Trials), cohort studies, case-control, case series, quasi-experimental studies etc...) | Study Protocols<br>Opinion pieces<br>Reviews |
| **OUTCOME MEASURES** | Technology acceptability data such as usability, user experience and/ effectiveness in crop farming | Absence of technology acceptability data in crop farming |

| STUDY ANALYSES | Qualitative analysis or statistical analysis of outcome data | No analysis of crop farming acceptability data reported |
|---|---|---|
| PUBLICATION CRITERIA | Human studies<br>Published in peer-reviewed journals and archival conference proceedings<br>Published in the English language | Published in a language other than English with an absence of peer-reviewed translation to English |

*Table 2: Keywords*

| **Agriculture** |
|---|
| agriculture* OR crop* OR farm* |
| **Technology-enabled Intervention** |
| "web base*" OR online OR "smart farm* OR internet OR comp* OR "smart agricult*" OR "IoT" OR "decision support" OR "machine learn*" OR "precision agricult" OR "visuali*" |
| **Acceptability** |
| accept* OR adopt* OR usab* |

As several of the databases did not support the merging of search results, screening was conducted separately for each database. Table 3 presents the screening results independently for each database. After merging the final results, a total of 63 articles remained after the initial screening process (see Table 3 for details). Each of these papers was read in detail and assessed by the first author RJT against the inclusion and exclusion criteria (see Table 1) of the study. All the 63 articles were re-examined by the fourth author DC of this paper. A few disparities were noted, and those were resolved by discussion and subsequent double-checking to ensure consistency. A total of 16 articles met the final criteria for inclusion. The example of the reasons articles from 63 cohort were excluded include (a) targeting a different population or user group, for example, users from the livestock farms (Fox et al. 2018; Michels, Hobe, and Musshoff 2020; Jarvis et al. 2017), (b) reported data is limited, for example where user evaluations were conducted but there is very limited or no results available (Yang, Jiao, and Zhang 2011) which cannot be verified as technology acceptance (Kaloxylos et al. 2014; Alemu and Negash 2015), (c) only usability is evaluated without any technology acceptance data (Sciarretta et al. 2019; Lasso et al. 2018; Castro et al. 2019), (d) only technology adoption is evaluated, for example, to understand agricultural practices (Sardar, Kiani, and Kuslu 2019), SMS technology (Beza et al. 2018) and socio-cultural factors (Arvila et al. 2018), (e) discussions on Smart Agriculture (SA) adoption, for example timing of technology adoption (Watcharaanantapong et al. 2014) and barriers of technology adoption (Aubert, Schroeder, and Grimaudo 2012), (f) agriculture technologies are evaluated in general, for example, intention and attitude toward adoption (Rezaei-Moghaddam and Salehi 2010; Tohidyan Far and Rezaei-Moghaddam 2017) and (h) intervention is developed but no user evaluation is conducted (Di Giovanni et al. 2012).

*Table 3: Paper selection process adapted to the PRISMA statement (Moher et al., 2009)*

| | ACM Digital Library | | Scopus | | CABI | | Business Source Complete | |
|---|---|---|---|---|---|---|---|---|
| **Identification** | Records identified through searching (n=1276) | | Records identified through searching (n=5781) | | Records identified through searching (n=2286) | | Records identified through searching (n=290) | |
| **Screening** | Records screened (n=1276) | Records excluded with reasons Title, Abstract and duplicates (n=944) | Records screened using database filters (n=5781) | Records excluded with reasons Not related to farming/ agriculture (crops) (n=2940) Background study/ models/ work in progress/ discussion/ review/ editorial papers (n=300) Not in English language (n=712) | Records screened using database filters (n=2286) | Records excluded with reasons Not related to farming/agriculture (crops) (n=509) Background study/ models/ work in progress/ discussion/ review/ editorial papers (n=532) Not in English language (n=276) | Records screened using database filters (n=290) | Records excluded with reasons Discussion/ editorial papers (n=79) Not related to farming/ agriculture (crops) (n=189) No full text available (n=19) |
| | | | Records screened (n=1829) | Records excluded with reasons Duplicates (n=201) | Records screened (n=969) | Records excluded with reasons Duplicates (n=164) | | |
| | | | Records screened (n=1628) | Records excluded with reasons Title and Abstract (n=1464) Retracted (n=6) | Records screened (n=805) | Records excluded with reasons Title and Abstract (n=712) | | |
| **Eligibility** | Full-text articles assessed for eligibility (n=332) | Full-text articles excluded with reasons Not within target population (n=5) Limited access to full text (n=26) DOI not found/not available online (n=39) Not related to farming/agriculture (crops) (n=10) Background study/ models/ work in progress/ discussion/ review papers (n=83) Absence of technology acceptance data/analysis in e-agriculture (n=138) Not directly used by farmers/agronomists (n=10) Not in English language (n=1) Retracted (n=2) | Full-text articles assessed for eligibility (n=158) | Full-text articles excluded with reasons Not within target population (n=4) Limited access to full text (n=5) DOI not found/not available online (n=5) Not related to farming/ agriculture (crops) (n=2) Background study/ models/ work in progress/ discussion/ review/ editorial papers (n=40) Absence of technology acceptance data/analysis in e-agriculture (n=69) Not directly used by farmers/agronomists (n=1) | Full-text articles assessed for eligibility (n=93) | Full-text articles excluded with reasons Limited access to full text (n=5) DOI not found/not available online (n=5) Not related to farming/agriculture (crops) (n=3) Background study/ models/ work in progress/ discussion/ review/ editorial papers (n=38) Absence of technology acceptance data/analysis in e-agriculture (n=29) | Full-text articles assessed for eligibility (n=3) | Full-text articles excluded with reasons Absence of technology acceptance data/analysis in e-agriculture (n=3) |
| **Included** | Articles included for detailed analysis (n=18) | | Articles included for detailed analysis (n=32) | | Articles included for detailed analysis (n=13) | | Articles included for detailed analysis (n=0) | |

Of the total of 16 articles that met the inclusion criteria, the research (n=2 each) was conducted in Australia, France, India and Indonesia. Others (n=1 each) were conducted in Brazil, China, Czech Republic, Germany, Italy, Philippines, Spain and Thailand. The articles were published between 2009 and 2020. Participant numbers ranged from 10 to 727, composed mainly of farmers and agronomists, but also including smaller groups such as lecturers, employees, researchers and students. One article (Jakku and Thorburn 2010), does not mention the number of participants. Quantitative and qualitative research methodologies were used across these research studies. Appendix A lists the authors of the 16 shortlisted articles along with the scale items.

The studies show that technology acceptance is described mainly in the terms of acceptability, adoption and even usability. A small number of studies differentiate these terms as concepts or as study hypotheses during the user participation for the technology acceptance study (pre-adoption, implementation, adoption stages etc...). However, most of the studies do not differentiate the stages but have an umbrella of constructs that effect technology acceptance.

The Theoretical Acceptance Model (TFA) illustrates acceptability from the user point of view using 7 constructs (Sekhon et al., 2017). They are (1) Affective Attitude which describes how users feel about the system (2) Burden which describes the perceived amount of effort needed by the user to use the system (3) Ethicality which describes the degree to which the system blends in with the user's value system. (4) Intervention/System Coherence which describes the degree to which the user understands the system and it's working (5) Opportunity Costs which captures what (and the degree to which), the user must be relinquish in order to use the system (6) Perceived Effectiveness describes the degree to the users perceive the system as delivering anticipated results (7) Self Efficacy which describes if the users are sufficiently confident to make the necessary behaviour changes required by the system. We mapped the shortlisted papers to the TFA as it would help us to capture all the elements (7 constructs) of technology acceptance. This can be clearly observed in Table 4 where the shortlisted articles mapped only with some of the constructs of TFA as it was only considered and/ evaluated as part of technology acceptance. Such a mapping can help us identify how technology acceptance is presently defined and captured in crop farming.

*Table 4: Mapping of 7 constructs of Theoretical Model of Acceptance to the shortlisted articles*

| Authors | AA | BD | ET | IC | OC | PE | SE |
|---|---|---|---|---|---|---|---|
| Jakku and Thorburn (2010) | X | | X | X | | X | |
| Li et al. (2020) | | | | | X | X | X |
| Iskandar et al. (2018) | X | | | X | | X | X |
| Mackrell et al. (2009) | | X | | X | | | X |
| Sayruamyat and Nadee (2020) | X | | | X | | X | |
| Ravier et al. (2018) | | X | | X | | | X |
| Souza et al. (2019) | | X | | X | | | X |
| del Aguila et al. (2015) | | | | X | | | X |
| Rahim et al. (2016) | X | X | | X | | | |
| Caffaro et al. (2020) | | | X | X | | X | |
| Jain et al. (2018) | X | | X | X | | X | X |
| Mir and Padma (2020) | X | | | X | | X | X |
| Ayerdi Gotor et al. (2020) | | | X | X | | X | X |
| Ulman et al. (2017) | X | | | X | X | X | |
| Michels et al. (2020) | | X | | X | | X | X |
| Mercurio and Hernandez (2020) | X | | | X | | X | X |

AA - Affective attitude, BD - Burden, ET - Ethicality, IC - Intervention coherence, OC - Opportunity costs, PE - Perceived Effectiveness & SE - Self Efficacy

It can be observed that Intervention/System Coherence (the degree to which the user understands the intervention and its working) was captured in the maximum number of 15 studies. This was followed by Perceived Effectiveness (the degree to the users perceives the intervention as deriving anticipated results) and Self Efficacy (if the users are confident enough to make the necessary behaviour changes required by the system) which were both captured in 11 studies. Therefore, it could be concluded that these 3 constructs of Intervention/System Coherence, Perceived Effectiveness and Self Efficacy were mainly (94% and 64%) considered as present evaluation criteria for technology acceptance in crop farming. However, this does not represent technology acceptance completely.

The remaining constructs were captured in a limited number of studies. Affective Attitude (how users feel about the system) was captured in 8 studies i.e., only 50% of the shortlisted studies. This is important as how the end-users perceive the technological intervention would effect the acceptance. Next, Burden (the perceived amount of effort needed by the user to use the intervention) and Ethicality (the degree to which the intervention blends in with the user's value system) which were captured in 5 and 4 studies respectively i.e., 31% and 25% of the shortlisted studies. Both, Burden and Ethicality influence technology acceptance as the End-users perceives the effort required to use the intervention and the extent to which it matches with their values. Finally, Opportunity Costs (the degree to which what must be relinquished by the users to use the system) was captured in the least number of 2 studies i.e., only 12.5% of the shortlisted studies which plays an equally important role in technology acceptance. This is going hand-in-hand with Burden but in a different context of 'user sacrifice' as they must forgo tangible and/ intangible elements to use the intervention. As a result, the present evaluations exclude important constructs on the whole that illustrate technology acceptance.

## Methodology

The results of the shortlisting of the literature are shown in the Appendix, which lists 198 scale items based on studies reported in these 16 papers. Unfortunately, most studies do not report on the scale construction, reliability, or validation. The exceptions are (Li et al. 2020; Caffaro et al. 2020; Mir and Padma 2020; Michels, Bonke, and Musshoff 2020), however, they all measured technology acceptance from different perspectives and have limitations.

Li et al. (2020) adapted the scale that was previously validated in the domain of mobile banking (Zhou, Lu, and Wang 2010) which produced cross-loadings and the items had to be removed but the authors' have not revalidated them in the farming domain. Caffaro et al. (2020) adapted items from scales validated in agriculture and information technology with a combination of adoption factors from agriculture (Davis 1989; Adrian, Norwood, and Mask 2005; Kernecker et al. 2019; Pierpaoli et al. 2013; Unay Gailhard, Bavorová, and Pirscher 2015), which gave rise to model fit issues and the authors' have not revalidated them in the farming domain. Michels, Bonke et al. (2020) adapted the scale items from information technology (Venkatesh et al. 2003). However, according to MacCallum et al. (1999), a scale with good internal consistency, at least three items per factor to have at least three or four items with high loadings per factor which is not followed by the adapted scale where one factor has only two items and another factor has only one item where internal consistency cannot be checked. Mir and Padma (2020) have not clearly explained where the scale items were derived from. The authors' provided the complete list of items (237) when contacted directly. These items were reduced to 105 after the reliability testing which is still not provided in the paper. However, the authors' have not followed (MacCallum et al. 1999) advice of having at least three items per factor to have at least three or four items with high loadings per factor as there were nine factors with only two items each. Their results suggest that there may be cross-loadings between their scale factors as they found a high correlation between a few factors which were not eliminated. This might explain why eight factors had at least five or more items. The authors' have not revalidated the 105 items scale in the farming domain.

We reduced the 198 items listed in the Appendix in four steps. First, RJT removed duplicates, merged highly similar items and grouped them into the 7 constructs. Next, the second author RW checked the items to either agree or disagree with RJT. The disagreements were resolved by further discussion. Second, RJT transformed items where possible so that it is applicable to assess farming systems or technologies in general (eg. items 14, 25, 30, 123, 149, 164 etc.). For instance, item 14 "PA may not improve the efficiency of agricultural management." was changed to "The system may not improve the efficiency of management.", item 25 "I accept the consolidation of small farms into cooperatives to adopt PA." was changed to "I accept the consolidation of my initiative into cooperatives to adopt the system.", item 30 "I have the necessary knowledge to use PA technologies." was changed to "I have the necessary knowledge to use the system.", item 149 "Extension experts who have been providing consultancy in IPM&NM for a long time feel threatened by IPM&NM-DSS." was changed to "The experts who have been providing consultancy for a long time feel threatened by the system.". Third, RJT and RW removed items for which this was not possible (items 1-5, 17-19, 23, 39, 47, 50, 51, 62, 65-69, 86, 91, 93, 97-102, 110, 114, 126-129, 131, 135, 136, 139, 140, 148, 152, 154, 167-170, 176, 183 and 194). This reduced the list to 57 items.

Furthermore, RJT and the third author AD had a discussion to further cross-check the items. Due to low or no relevance in the actual farming domain, we removed some more items (item [20: 193, 81, 21], 25, 55, [79: 77, 83, 118, 107, 108], [106: 174], [122: 119, 120, 121], [182: 70, 71]). The sub-items were the 'derived from' items for the main item. Next, we transformed the language of some items to make them easy to understand (eg. item 96 "I enjoy using the system with an attractive user interface." was changed into "I enjoy using the system with attractive graphics.", item 143 "We possess those characteristics which a user of the system is supposed to have." was changed to "This system was made for farmers like me.", item 13 "The system may increase the costs and may not improve the revenue." as changed to "The system may not show a positive cost to benefit ratio." etc. Then we combined a few more similar items (item 105 with 155, 7 with 34, and 146 with 145). This reduced the list to 47 items used for the initial scale development as shown in Table 5, which also shows which original items these were derived from. Finally, RJT and the fourth author DC had a discussion to the cross-check and correct the items for grammatical mistakes.

*Table 5. Scale items developed*

| No. | Scale Item | Duplicate/Similar/Derived From |
|---|---|---|
| | **AA Affective Attitude which describes how users feel about the system** | |
| 96 | I enjoy using the system because of good graphics. | 82, 72 |
| 111 | If I heard that a new version were available, I would be interested in testing it. | |
| 112 | I prefer to use the most advanced system available. | |
| 113 | In general, I hesitate to try a new system. | |
| 161 | I think the system is useful and advantageous. | 162 |
| 173 | I do not intend to use this system anymore in the future. | |
| | **BD Burden which describes the perceived amount of effort needed by the user to use the system** | |
| 40 | I can use the system with little effort. | 43 |
| 84 | It is difficult to learn to use the system. | |
| 130 | Education and training are a must and provide confidence to use the system. | 133 |
| 32 | If I have any doubts about how to use the system there will be professionals to help me train. | 132, 85 |
| 198 | My smartphone and mobile internet coverage are sufficient to use the system. | |
| | **ET Ethicality which describes the degree to which the system blends in with the user's value system** | |
| 123 | The system is well suited for our farm business practices. | 124, 137 |
| 125 | We support the introduction of this new system that will help management. | |
| 138 | I like using the system because it fits with my personal ethos of farming. | 134 |
| 143 | This system was made for farmers like me. | 142 |
| 150 | Resistance to change could be a problem for adopting this system. | 153 |
| | **IC Intervention/System Coherence which describes the degree to which the user understands the system and it's working** | |
| 37 | The user interface is easy to use, understand, intuitive and well suited to the device. | 166, 38, 41, 179, 184 |
| 78 | The system is simple, easy to learn and control. | 90, 185, 190, 163, 191, 192, 165 |
| 42 | The system options are relevant. | |
| 46 | The system provides error-free and adequately explained responses. | 48, 49, 53, 60, 61 |
| 59 | The system is easy to install, learn, input data and output is useful. | 56, 57, 58 |
| 141 | We understand the advantages offered by this system over others. | |
| 164 | I can easily access all the information I need. | 52 |
| | **OC Opportunity Costs which captures what (and the degree to which), the user must be relinquish in order to use the system** | |
| 13 | The system may not have a positive cost to benefit ratio. | 15 |
| 14 | The system may not improve the efficiency of management. | |
| 16 | The system may have technical risks due to its uncertainty. | 151 |
| 24 | I am willing to buy the system as I have access to finance or credit. | 29 |
| 92 | The system protects user privacy and is risk-free. | 95 |
| 149 | People who advise farmers will feel threatened by this system. | |
| | **PE Perceived Effectiveness describes the degree to the users perceive the system as delivering anticipated results** | |
| 76 | The system enables me to work more efficiently. | 177, 63, 88, 189, 44, 11 |
| 155 | The system can improve my decision-making efficiency. | 87, 103, 104, 105, 94 |
| 157 | The system improves my knowledge. | |
| 34 | The system could reduce overall costs by keeping track of resources and making recommendations. | 158, 7, 12, 27, 33, 156, 159, 6, 8, 26, 28, 10, 35, 187 |
| 73 | The system is useful for farm operations as it increases productivity and lowers production costs. | 74, 75 |
| 175 | The system would provide big advantages to the users. | |
| 186 | The system would be useful in my day-to-day work and meet my expectations for task completion. | 9, 181, 180, 178, 64 |
| 9 | The system could meet my needs. | |

| 188 | The system could reduce activities that have negative impacts on the environment. | 36 |
|---|---|---|
| 160 | The system can improve communication within discussion groups. | |
| | **SE Self Efficacy which describes if the users are sufficiently confident to make the necessary behaviour changes required by the system** | |
| 30 | I have the necessary knowledge to use the system. | |
| 31 | I have the necessary resources to use the system. | |
| 45 | I would use this system. | |
| 54 | I think the system can do everything I need. | |
| 115 | It would be easy to use and adopt this system. | 116, 117, 109 |
| 145 | I use the system as often as possible. | 144, 146 |
| 171 | I intend to use the system frequently in future as well. | 147, 195, 196, 197 |
| 172 | I intend to use the system occasionally in the future. | 22, 89 |

The developed scale of 47 shortlisted items needs to be validated with a user study to be used as a standardised scale in smart agriculture. The data from the study of the standardised scale will be analysed using Exploratory Factor Analysis (Principal Component Analysis -PCA- extraction) to cross check if the items belong to specified construct and if there are correlations. For the factor analysis (PCA), all the 5-point scale items will be considered as ordinal measures. This will be then followed by the Confirmatory Factor Analysis to determine the validity of the scale, and to confirm the constructs/factors. This will help in the cross-validation of the items across the scale and in the reduction of the total items. The validated Technology Acceptance Scale can be used to measure acceptance for an existing product. It can be also used in the co-creation or development of new tools and products in smart agriculture.

### Limitations
The development of the scale was based on the literature screening on crop-based agriculture that focused specifically on intelligent technologies such as decision support systems without considering acceptance of broader technology in agriculture. We acknowledge the possibility of subjective judgement in shortlisting articles and the merging of the scale items. Additionally, the mapping of the shortlisted articles to the Theoretical Model of Acceptance is a subjective representation from the literature analysis. Therefore, these limitations may not necessarily transfer to the developed scale as it is a combination of all the collected scale items. However, this can be confirmed only after the data collection process when the scale will be implemented.

### Conclusions and future work
In this paper, we developed a shortlisted scale of 47 items, which can be used to evaluate technology acceptance in crop farming. However, we need to validate the scale and produce a shortlisted validated version for future use. The formulation of such a scale would contribute to closing the gap that exists in the present non-validated methods of technology acceptance evaluations in this domain. Future work across the community on a validated scale would allow us to frame research questions within technology acceptance constructs, allowing for the validation of a standard scale in crop farming and possibly the wider agricultural community.

# Appendix A

*Table 6. Scale items from the literature review*

| Authors | Type | No | Scale Item |
|---|---|---|---|
| (Li et al. 2020) | Quantitative | 1 | I need to improve working efficiency on farming. |
| | | 2 | I need to optimize the level of fertilizer that I use. |
| | | 3 | I need to optimize the level of pesticide that I use. |
| | | 4 | I need to optimize the amount of seeds that I use. |
| | | 5 | I need to optimize the amount of water that I use. |
| | | 6 | PA allows me to map the location of problems with crops. |
| | | 7 | PA provides useful information about crop growing. |
| | | 8 | PA provides useful information for farm decisions. |
| | | 9 | PA technologies can meet my needs. |
| | | 10 | PA allows me to manage my farm more efficiently. |
| | | 11 | PA allows me to manage my farm more easily. |
| | | 12 | PA reduces water usage. |
| | | 13 | PA may increase the costs of farming. |
| | | 14 | PA may not improve the efficiency of agricultural management. |
| | | 15 | PA may not improve the grower's revenue. |
| | | 16 | PA may have technical risk. |
| | | 17 | Local communities encourage me to adopt PA on my farm. |
| | | 18 | The Government encourages me to adopt PA on my farm. |
| | | 19 | Cooperatives encourage me to adopt PA on my farm. |
| | | 20 | PA helps me get a high positive profile amongst other farmers. |
| | | 21 | PA helps me to get certification for New type professional farmer. |
| | | 22 | I would like to adopt PA in the future. |
| | | 23 | I would like to be a service provider for PA. |
| | | 24 | I am willing to pay for a PA service |
| | | 25 | I accept the consolidation of small farms into cooperatives to adopt PA. |
| | | 26 | PA improves soil condition. |
| | | 27 | PA reduces pesticide residues. |
| | | 28 | PA increases product quality. |
| | | 29 | I have access to the financial support I need to use PA. |
| | | 30 | I have the necessary knowledge to use PA technologies. |
| | | 31 | I have the necessary resources to use PA technologies. |
| | | 32 | If I have any doubts about how to use the PA, there will be professionals to help me. |
| | | 33 | PA reduces investments in seeds. |
| | | 34 | PA reduces investments in fertilizers. |
| | | 35 | PA reduces investments in pesticides. |
| | | 36 | PA reduces the negative impacts of agricultural activities on environment. |
| (Souza et al. 2019) | Quantitative | 37 | The user interface is easy to understand. |
| | | 38 | The user interface is easy to use. |
| | | 39 | The option are clear and objectives |
| | | 40 | With little effort I can select a context of interest. |
| | | 41 | The user interface is properly adapted to the devices. |
| | | 42 | The options presented are relevant. |
| | | 43 | The solution makes it easy to obtain contextual data from multiple sensors. |
| | | 44 | The solution facilitates immediate action from the issuance of an alert or message. |
| | | 45 | I would use this solution in my work? |
| (del Águila, Cañadas, and Túnez 2015) | Both | 46 | Does the system adequately explain why it has produced a specific response? |
| | | 47 | Does the system justify why certain information has been requested? |

| | | 48 | Are the explanation messages adequate? |
|---|---|---|---|
| | | 49 | Does the system adequately explain the special situations? |
| | | 50 | Is the information contained organized logically? Does it highlight important information? |
| | | 51 | Do you know exactly how and where to introduce new information? |
| | | 52 | When a task is performed with the system, do you think you can access all the information you need? |
| | | 53 | Does the system help you with the appropriate information (error messages, warnings, help information, etc.)? |
| | | 54 | Do you think the system can do everything you need? |
| | | 55 | Does the system respond appropriately to user actions? |
| (Rahim, Supli, and Damiri 2016) | Quantitative | 56 | Ease to input the data |
| | | 57 | Ease to use output |
| | | 58 | Ease to install |
| | | 59 | Ease to learn |
| | | 60 | Accurate |
| | | 61 | Should be error-free |
| | | 62 | Successful |
| | | 63 | Time taken o respond |
| | | 64 | Time taken o complete a task |
| | | 65 | Support/help |
| | | 66 | Touch screen facilities |
| | | 67 | Voice guidance |
| | | 68 | System resources info |
| | | 69 | Automatic update |
| | | 70 | Safety while using the application |
| | | 71 | Safety while driving |

|---|---|---|---|
| | | 72 | Attractiveness of user interface |
| (Caffaro et al. 2020) | Quantitative | 73 | it is useful for farm operations |
| | | 74 | it increases productivity |
| | | 75 | it lowers production costs |
| | | 76 | it allows one to work more quickly; it reduces workload |
| | | 77 | it is easy to learn; it is controllable |
| | | 78 | it is understandable |
| | | 79 | it is easy to become skilful in. |
| | | 80 | whether they were going to adopt the chosen SFT during the following year? |
| | | 81 | how often they were exposed to different sources of information about the chosen SFT, such as trade journals, advertisements, training courses, internet/social media, exhibitions, and discussions with peers, consultants, trade organisations, and relatives? |
| (Jain et al. 2018) | Both | 82 | Enjoyment |
| | | 83 | Easy to Use |
| | | 84 | Difficult to Learn |
| | | 85 | Need Support |
| | | 86 | Understands Me |
| | | 87 | Trust Response |
| | | 88 | Quick Response |
| | | 89 | Will Use in Future |
| (Mir and Padma 2020) | Both | 90 | I have a clear conception of the functionalities of the IPM&NM-DSS. |
| | | 91 | The IPM&NM-DSS developer is widely acknowledged. |
| | | 92 | IPM&NM-DSS protects privacy of its users. |
| | | 93 | I feel confident that I can keep IPM&NM-DSS under my own control. |
| | | 94 | I feel the data returned by the system is reliable. |

| | |
|---|---|
| 95 | I feel it is risk free to use IPM&NM-DSS. |
| 96 | I enjoy using new technologies like DSS. |
| 97 | New technologies like DSS make my life easier. |
| 98 | I feel like I am overly dependent on technology. |
| 99 | Technology allows me to do the things more easily. |
| 100 | I think DSS developers convince us that we need things that we don't really need |
| 101 | The more I use a new technology, the more I become a slave to it. |
| 102 | New technology makes it too easy for companies and other people to invade my privacy |
| 103 | The IPM&NM-DSS will offer reliable information. |
| 104 | I expect the quality of the decision support obtained from the IPM&NM-DSS will be of high quality. |
| 105 | The quality of decision support by the IPM&NM-DSS is correct or close to the true value. |
| 106 | Accuracy is maintained throughout in the IPM&NM-DSS. |
| 107 | I could get a decision support using apple IPM&NM-DSS if I had never used a system like it before. |
| 108 | I could complete the decision support process using apple IPM&NM-DSS if I had used similar system before |
| 109 | I have the ability to operate apple IPM&NM-DSS. |
| 110 | I prefer to use an IPM&NM-DSS for extension purpose. |
| 111 | If I heard that a new DSS is available, I would be interested enough to test. |
| 112 | I prefer to use the most advanced DSS available |
| 113 | In general, I hesitate to try new DSS system. |
| 114 | There are many farmers and extension specialists, who have already adopted some DSSs. |
| 115 | We have sufficient expertise to adopt DSS technology. |
| 116 | It would be easy for us to adopt DSS technology. |
| 117 | We have the necessary skills and capabilities required to succeed on the DSS technology paradigm. |
| 118 | IPM&NM-DSS is designed for the user with particular education levels. |
| 119 | IPM&NM-DSS is more important for user with higher education levels. |
| 120 | More users having higher education levels have the ability to use IPM&NM-DSS. |
| 121 | Working with IPM&NM-DSS is more for users having higher education levels |
| 122 | Users having higher education levels can do just as well as the users having middle levels of education while using IPM&NM-DSS. |
| 123 | AIP&NM-DSS is meaningful and has relevance in IPM and INM practices. |
| 124 | Farmers and extension specialists are relevant users of AIP&NM-DSS. |
| 125 | We support introduction of new technology for the management of orchards |
| 126 | We update H/W, S/W and application software's regularly. |
| 127 | Using IMP&NM-DSS, I/we have exploited more intervention opportunities from its main competitors |
| 128 | Using IMP&NM-DSS, I/we have not exploited any of the intervention opportunities from its main competitors. |
| 129 | Using IMP&NM-DSS, I/we have neutralized most competitive threats from our main competitors. |
| 130 | Education and training is must for using IPM&NM-DSS. |
| 131 | There exists a policy to educate and train end-user groups. |
| 132 | Every IPM&NM-DSS user was properly trained |
| 133 | The education and training provide us confidence in the use of IPM&NM-DSS |
| 134 | What IPM&NM-DSS stands for is important for me as a stakeholder. |
| 135 | In order to prepare for my career in apple fruit production, it is necessary to learn apple IPM&NM-DSS. |
| 136 | We are actively encouraged to pursue our own ideas related to IPM&NM-DSS adoption. |
| 137 | There are many farmers and extension specialists having good ideas to start using IPM&NM-DSS. |
| 138 | I like using apple IPM&NM-DSS based on the similarity of my values and society values underlying its use. |
| 139 | I like to recommend IPM&NM-DSS to my cosmopolitans. |
| 140 | We are aware of DSS implementation by our competitors. |
| 141 | We understand the competitive advantages offered by IPM&NM-DSS. |
| 142 | IPM&NM-DSS has been developed for average user characteristics. |
| 143 | We possess those characteristics which a user of IPM&NM-DSS is supposed to have. |
| 144 | I always try to use AIP&NM-DSS for decision support whenever it has a capability and feature. |
| 145 | I always try to use AIP&NM-DSS as many cases as possible. |
| 146 | I expect to use AIP&NM given its accessibility. |
| 147 | I intend to use AIP&NM-DSS continuously in the future. |
| 148 | Behavioural changes in our organization are very slow |
| 149 | Extension experts who have been providing consultancy in IPM&NM for a long time feel threatened by IPM&NM-DSS. |
| 150 | Although it seems that IPM&NM-DSS brings the change, in practice it has not. |
| 151 | I are afraid because of the uncertainty in the decision support by the new way of working with IPM&NM-DSS |
| 152 | Change generates opportunities for those who know how to take advantage of IPM&NM-DSS. |
| 153 | Farmers tend to pretend they agree with the changes, but in reality do not allow the changes to be introduced. |
| 154 | People are reluctant to try to understand the fundamental objectives of the changes |
| 155 | IPM&NM-DSS can improve my decision making efficiency. |
| 156 | IPM&NM-DSS is helpful in the assessment of disease, pests and nutrient imbalances and control provisions. |
| 157 | IPM&NM-DSS improves my knowledge about IPM and INM. |
| 158 | The IPM&NM-DSS would make it easier to keep track of spray schedules and fertilizer applications. |
| 159 | IPM&NM-DSS is helpful to keep track of vulnerabilities in the weather conditions. |
| 160 | Using apple IPM&NM-DSS can improve communication between farmers and extension functionaries |
| 161 | I think that IPM&NM-DSS is useful. |
| 162 | Overall, apple IPMDSS is advantageous. |
| 163 | Learning to use IPM&NM-DSS would be easy for me. |
| 164 | I would easily get the information; I am looking for using IPM&NM-DSS. |

| | | | |
|---|---|---|---|
| | | 165 | It would be easy for me to become skilful at using IPM&NM-DSS. |
| | | 166 | The user interface of IPM&NM-DSS is easy and intuitive. |
| | | 167 | English language is not a barrier when I use IPM&NM-DSS because other language are also available. |
| | | 168 | I am using IPM&NM-DSS once in every 3 months. |
| | | 169 | I am using IPM&NM-DSS more frequently. |
| | | 170 | I am not using IPM&NM-DSS at all |
| | | 171 | I intend to use IPM&NM frequently in future as well. |
| | | 172 | I intend to use IPM&NM occasionally in the future. |
| | | 173 | I do not intend to use IPM&NM any more in the future. |
| (Ulman et al. 2017) | Quantitative | 174 | How accurately the service works? |
| | | 175 | How big advantages are brought to you by the service? |
| | | 176 | How often do you use the service? |
| | | 177 | How fast response does the service provide? |
| | | 178 | How well does the service meet your requirements? |
| | | 179 | To what extent is the service web site user friendly? |
| | | 180 | How are you satisfied with service functions? |
| | | 181 | Does the service meet your expectations? |
| | | 182 | To what extent is the use of service safe for you? |
| | | 183 | How well are information organized on the website? |
| | | 184 | How easy is to navigate on the page? |
| | | 185 | To what extent does the page help you to understand its control? |
| (Michels, Bonke, and Musshoff 2020) | Quantitative | 186 | I would find the use of crop protection smartphone apps useful in my day-to-day work |
| | | 187 | I think using crop protection smartphone apps would make my crop protection more cost-effective |
| | | 188 | I think crop protection smartphone apps would make crop protection more environmentally friendly |
| | | 189 | I think that crop protection smartphone apps would speed up my work completion |
| | | 190 | The handling of a crop protection smartphone app would be simple and understandable for me |
| | | 191 | All in all, I would find crop protection smartphone apps to be easy to use tools |
| | | 192 | I believe that the use of crop protection smartphone apps would be easy for me to learn |
| | | 193 | Farmers who are friends of mine think that it makes sense to use a crop protection smartphone app |
| | | 194 | My work colleagues think that I should use crop protection smartphone apps |
| | | 195 | I plan to use crop protection smartphone apps |
| | | 196 | I intend to use crop protection smartphone apps |
| | | 197 | It is likely that I will use crop protection smartphone apps in the future |
| | | 198 | My smartphone and mobile internet coverage are sufficient to use a crop protection smartphone app |
| (Mercurio and Hernandez 2020) | Quantitative | | No access to items used, emailed authors |
| (Iskandar, Rosmansyah, and Albarda 2018) | Quantitative | | No access to items used, emailed authors |
| (Mackrell, Kerr, and von Hellens 2009) | Qualitative | | |
| (Sayruamyat and Nadee 2020) | Quantitative | | No access to items used, emailed authors |
| (Ravier et al. 2018) | Qualitative | | |
| (Jakku and Thorburn 2010) | Qualitative | | |
| (Ayerdi Gotor et al. 2020) | Qualitative | | |